\documentclass[12pt,english]{article}
\usepackage[T1]{fontenc}
\usepackage{geometry}
\usepackage{amsmath}
\usepackage{babel}
\usepackage{graphics}
\usepackage{amssymb}

\makeatletter

\providecommand{\LyX}{L\kern-.1667em\lower.25em\hbox{Y}\kern-.125emX\@}

\makeatother
\begin{document}

\title{Macroscopic entanglement jumps in model spin systems}

\author{Indrani Bose and Emily Chattopadhyay}

\maketitle
{\centering \textbf{Physics Department, Bose Institute, 93/1, A.P.C.
Road, Calcutta-700009, India}\par}

\begin{abstract}
In this paper, we consider some frustrated spin models for which the
ground states are known exactly. The concurrence, a measure of the
amount of entanglement can be calculated exactly for entangled spin
pairs. Quantum phase transitions involving macroscopic magnetization
changes at critical values of the magnetic field are accompanied by
macroscopic jumps in the \( T=0 \) entanglement. A specific example
is given in which magnetization plateaus give rise to a plateau structure
in the amount of entanglement associated with nearest-neighbour bonds.
We further show that macroscopic entanglement changes can occur in
quantum phase transitions brought about by the tuning of exchange
interaction strengths. 
\end{abstract}
Entanglement is a characteristic feature of quantum mechanical systems
which has no classical analogue \cite{key-1}. The state of a pair
(or more than a pair) of quantum systems is entangled if the corresponding
wave function does not factorize, i.e., is not a product of the wavefunctions
of the individual systems. A well-known example of an entangled state
is the singlet state of two spin\( -\frac{1}{2} \) particles, \( \frac{1}{\sqrt{2}}\left( \left| \uparrow \downarrow \right\rangle -\left| \downarrow \uparrow \right\rangle \right)  \),
which cannot be written as a product of the spin states of individual
spins. Measurement on one component of an entangled pair fixes the
state of the other implying non-local correlations. Interest in quantum
entanglement is extensive because of its fundamental role in quantum
communication and information processing such as quantum teleportation
\cite{key-2}, superdense coding \cite{key-3}, quantum cryptographic
key distribution \cite{key-4} etc. Experimental implementation of
some of the protocols has so far been achieved in simple physical
settings. Solid state devices, specially, spin systems have been proposed
as possible candidates for large scale realizations \cite{key-5,key-6}.
In particular, the Heisenberg spin-spin exchange interaction gives
rise to entangled states in spin systems and has been shown to provide
the basis for universal quantum computation \cite{key-7,key-8}. Examples
of other interacting many body systems in which entanglement properties
have been studied include the harmonic chain \cite{plenio}, the 1D
Kondo Necklace model \cite{saguia} and the BCS condensate \cite{martin}.

Entanglement in a state like its energy is quantifiable and has been
computed both at \( T=0 \) and at finite \( T \) (thermal entanglement)
for a variety of spin models in both zero and finite external magnetic
fields. The models include the Heisenberg \( XX \), \( XY \), \( XXX \),
\( XXZ \) and transverse Ising models in one dimension (1D) {[}12-18{]}.
The computational studies show that the amount of entanglement between
two spins in a multi-spin state can be modified by changing the temperature
and/or the external magnetic field. Some recent studies have explored
the relations between entanglement and quantum phase transitions in
the ferromagnetic (FM) \( XY \) model in a transverse magnetic field
and in a special case of the model, namely, the transverse Ising model
\cite{key-13,key-14}. A quantum phase transition (QPT) can take place
at \( T=0 \) by changing some parameter of the system or an external
variable like the magnetic field \cite{key-15}. The ground state
wavefunction undergoes qualitative changes in a QPT and model studies
indicate that entanglement develops special features in the vicinity
of a quantum critical point.

The studies carried out so far have been confined to 1D systems with
only nearest-neighbour (NN) exchange interactions. There are several
quasi-1D and 2D antiferromagnetic (AFM) spin models which describe
frustrated spin systems \cite{bose}. Frustration arises if conflicting
interactions are present in the system, i.e., when all the interactions
between spins cannot be simultaneously satisfied. A good example of
a frustrated spin system is the AFM Ising model on the triangular
lattice. An elementary plaquette of the lattice is a triangle. The
Ising spin variables have two possible values, \( \pm 1 \), corresponding
to the up and down spin orientations. An antiparallel spin pair has
the lowest interaction energy \( -J \). A parallel spin pair has
the energy \( +J \). In an elementary triangular plaquette, there
are three interacting spin pairs. Due to the topology of the plaquette
(odd number of NN bonds), all the three spin pairs cannot be simultaneously
antiparallel in the ground state. There is bound to be one parallel
spin pair giving rise to frustration in the system. Consider another
example in which the spins in a linear chain interact via NN as well
as next-nearest-neighbour (NNN) AFM interactions. In a triplet of
successive spins, the two NN spin pairs and one NNN spin pair cannot
be simultaneously antiparallel and the linear spin chain is frustrated.
Frustration in a system may occur due to the topology of the lattice
(triangular, kagom\'{e} etc.) or due to the inclusion of further-neighbour
interactions leading in most cases to spin-disordered ground states.
The models exhibit QPTs as the exchange interaction parameters are
tuned to particular ratios or at critical values of the external magnetic
field. Some of the models exhibit the phenomenon of magnetization
plateaus in which plateaus appear in the magnetization/site \( m \)
versus the external magnetic field \( h \) curve at quantized values
of \( m \) (\emph{m} and \( h \) are chosen to be dimensionless)
\cite{key-16}. The condition for the appearance of plateaus is 

\begin{equation}
\label{1}
S_{U}-m_{U}=\textrm{integer}
\end{equation}
 where \( S_{U} \) and \( m_{U} \) are the total spin and magnetization
in unit period of the ground state. The plateaus indicate that the
spin excitation spectrum is gapped so that the magnetization remains
constant. In this paper, we consider some spin models for which the
ground states are known exactly in certain parameter regimes. An exact
measure of the entanglement between two spins can be obtained in these
states in a straightforward manner because of the simple structure
of the ground states. We report two major results. In the presence
of an external magnetic field, macroscopic magnetization jumps at
critical values of the magnetic fields (first order QPTs) are accompanied
by macroscopic changes in the amount of pairwise entanglement. Furthermore,
some examples are given in which the entanglement structure is modified
due to QPTs (again first order) brought about by the tuning of exchange
interaction strengths. 

A measure of entanglement between the spins \( A \) and \( B \)
is given by a quantity called concurrence \cite{key-9,key-10}. To
calculate this, a knowledge of the reduced density matrix \( \rho _{AB} \)
is required. This is obtained from the ground state wavefunction by
tracing out all the spin degrees of freedom except those of the spins
\( A \) and \( B \). Let \( \rho _{AB} \) be defined as a matrix
in the standard basis \( \left\{ \left| \uparrow \uparrow \right\rangle ,\left| \uparrow \downarrow \right\rangle ,\left| \downarrow \uparrow \right\rangle ,\left| \downarrow \downarrow \right\rangle \right\}  \).
One can define the spin-reversed density matrix as \( \widetilde{\rho }=\left( \sigma _{y}\otimes \sigma _{y}\right) \rho ^{*}\left( \sigma _{y}\otimes \sigma _{y}\right)  \),
where \( \sigma _{y} \) is the Pauli matrix. The concurrence \( C \)
is given by \( C=\textrm{max}\{\lambda _{1}-\lambda _{2}-\lambda _{3}-\lambda _{4},0\} \)
where \( \lambda _{i}'s \) are the square roots of the eigenvalues
of the matrix \( \rho \widetilde{\rho } \) in descending order. The
spins \( A \) and \( B \) are entangled if \( C \) is nonzero,
\( C=0 \) implies an unentangled state and \( C=1 \) corresponds
to maximum entanglement. The models we consider in this paper belong
to the Majumdar-Ghosh (MG) \cite{key-17} and the Affleck-Kennedy-Lieb-Tasaki
(AKLT) \cite{key-18} families of models. The MG model is the simplest
frustrated model in 1D. The spins have magnitude \( \frac{1}{2} \)
and the Hamiltonian is given by

\begin{equation}
\label{2}
H_{MG}=J\sum ^{N}_{i=1}\overrightarrow{S}_{i}.\overrightarrow{S}_{i+1}+\frac{J}{2}\sum ^{N}_{i=1}\overrightarrow{S}_{i}.\overrightarrow{S}_{i+2.}
\end{equation}
 \( N \) is the total number of spins and the strength of the next-nearest-neighbour
(NNN) exchange interaction is half that of the NN interaction. The
boundary condition is periodic. The exact ground state is doubly degenerate
with the wave functions

\begin{eqnarray}
\phi _{1} & = & [12][34].......[N-1N]\nonumber \\
\phi _{2} & = & [23][45]........[N1]\label{3} 
\end{eqnarray}
where \( [lm] \) denotes a singlet (valence bond (VB)) of two spins
located at the lattice sites \( l \) and \( m \). Thus in the ground
state NN spin pairs in singlet spin configurations are maximally entangled
with concurrence \( C=1 \). The value of \( C \) for all the other
NN or further-neighbour pairs is zero. This is consistent with the
special property of entanglement that in a set of three spins \( A \),
\( B \) and \( C \), if \( A \) and \( B \) are maximally entangled,
the entanglement between \( A \) and \( C \) is zero \cite{key-9}.
Translational invariance in the ground states can be restored by taking
linear combinations of the ground states, \( \frac{\phi _{1}\pm \phi _{2}}{\sqrt{2}} \).
In this case, all NN pairs are entangled with \( C=0.5 \). The value
of \( C \) in the case of the \( S=\frac{1}{2} \) Heisenberg AFM
spin chain is \( C=0.386 \) \cite{key-9}. Frustration appears to
increase the NN entanglement in a spin system.
\begin{figure}[ihtbp]
{\centering \resizebox*{4in}{!}{\includegraphics{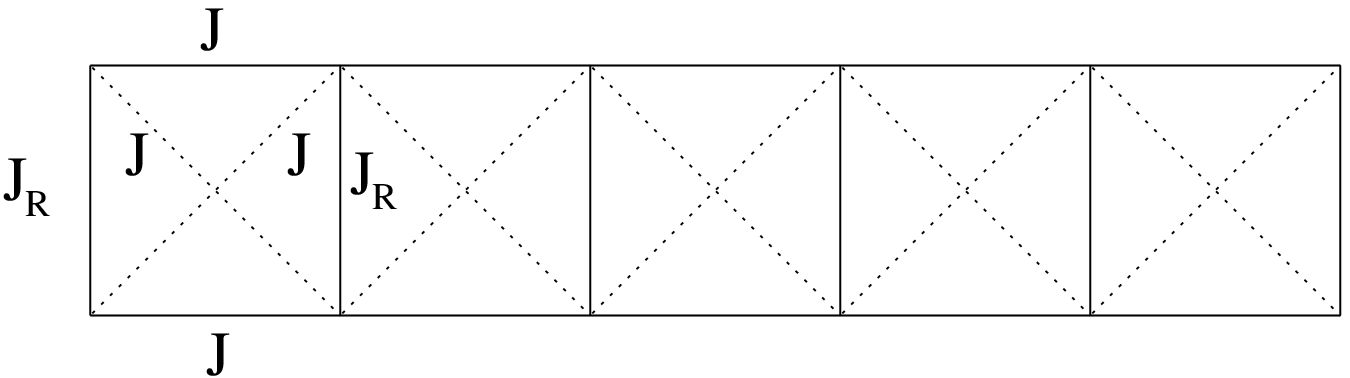}} \par}

{\centering FIG. 1: A two-leg frustrated spin ladder. The spin-spin
exchange interaction strength along the rung is \( J_{R} \). The
inter-leg NN and the diagonal exchange interactions are of equal strength
\( J \).\par}
\end{figure}

We now give examples of some spin models in the presence of an external
magnetic field for which macroscopic magnetization jumps are accompanied
by macroscopic jumps in entanglement. The first model is a frustrated
two-leg \( S=\frac{1}{2} \) ladder model \cite{key-19} (Fig. 1)
the Hamiltonian of which is given by

\begin{equation}
\label{4}
H_{ladder}=\sum _{\left\langle ij\right\rangle }J_{ij}\overrightarrow{S}_{i}.\overrightarrow{S}_{j}-h\sum ^{N}_{i=1}S^{z}_{i}.
\end{equation}
The exchange interaction strength \( J_{ij}=J_{R} \) along the rungs.
The intra-chain NN and the diagonal exchange interactions are of equal
strength \( J \). When \( h=0 \) and \( \frac{J_{R}}{J}>\lambda  \)
\( (\lambda \simeq 1.401) \), the exact ground state consists of
singlets (VBs) along the rungs. The spin pairs along the rungs are
thus maximally entangled and all other spin pairs are unentangled.
We define a quantity 

\begin{equation}
\label{5}
f=\frac{2}{N}\sum ^{\frac{N}{2}}_{i=1}C(i)
\end{equation}
 which is the average concurrence per rung of the ladder. The summation
is over the rung index with \( \frac{N}{2} \) being the total number
of rungs. Since all the other spin pairs are unentangled, the average
concurrence per NN bond, \( f_{NN}=\frac{1}{3} \) (the NN bonds include
\( \frac{N}{2} \) rung bonds and \( N \) intra-chain NN bonds).
The magnetization properties of the frustrated ladder model are simple
\cite{key-20}.
\begin{figure}[ihtbp]
{\centering \resizebox*{4.5in}{!}{\includegraphics{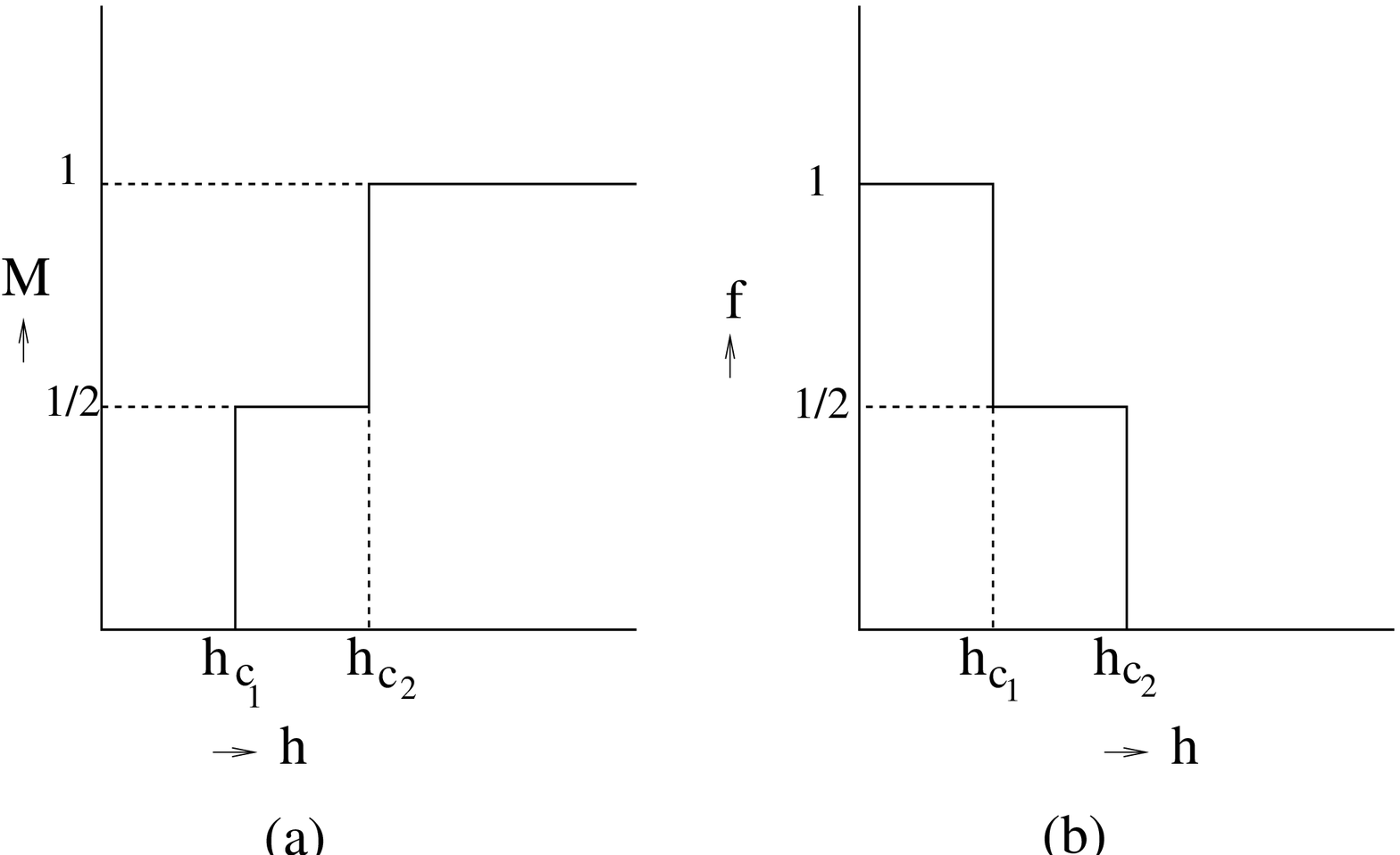}} \par}

{\centering FIG. 2: The frustrated spin-ladder model (Fig. 1) exhibits
plateaus in (a) magnetization/rung \emph{M} versus an external magnetic
field \emph{h} and (b) average concurrence per rung \emph{f} versus
\emph{h}.\par}
\end{figure}
 Fig. 2(a) shows the magnetization per rung \( M \) as a function
of the magnetic field \( h \). For \( 0<h<h_{c_{1}}=J_{R} \), \( M=0 \).
For \( h_{c_{1}}<h<h_{c_{2}}=J_{R}+2J \), the rungs are alternately
in singlet and \( S^{z}=1 \) triplet spin configurations in the ground
state. The value of \( M \) is now \( \frac{1}{2} \). For \( h>h_{c_{2}} \),
saturation magnetization is obtained with \( M=1 \). One can verify
that the quantization condition (1) is obeyed at each plateau. Fig.
2(b) shows the plot of the average concurrence per rung \( f \) versus
the magnetic field \( h \). A similar plot is obtained for \( f_{NN} \)
with the difference that \( f_{NN}=\frac{1}{3} \) for \( 0<h<h_{c_{1}} \)
and \( f_{NN}=\frac{1}{6} \) for \( h_{c_{1}}<h<h_{c_{2}} \). For
\( h_{c_{1}}<h<h_{c_{2}} \), half the total number of rungs are in
\( S^{z}=1 \) triplet spin configurations in which the two spins
of a rung point up. The parallel spin pairs are unentangled with \( C=0 \).
The other rungs are in singlet spin configurations so that \( f=\frac{1}{2} \).
In the fully polarised state \( f=0 \). Fig. 2(b) provides an example
of a changing magnetic field giving rise to macroscopic changes in
entanglement. The ground state of the frustrated spin ladder model
for \( h=0 \) has a simple structure and can be expressed as a product
over rung singlet states. The reduced density matrix \( \rho _{AB} \),
needed to calculate the entanglement between the spins \( A \) and
\( B \), can be calculated exactly and analytically in a straightforward
manner. Once the matrix elements of \( \rho _{AB} \) are known, the
concurrence \( C \) can be calculated. Because of the uncomplicated
structure of the ground state, \( C \) can alternatively be calculated
using simple arguments. Each spin in the ladder belongs to a rung
singlet or VB so that it is maximally entangled with its partner spin.
Thus, the entanglement between a spin \emph{j} and any other spin,
belonging to other rung singlets, is zero. The concurrence \( C \)
of a maximally entangled pair is 1 so that \( f, \) the average concurrence
per rung, is 1 for \( 0<h<h_{c_{1}}. \) Similar arguments hold true
for \( h_{c_{1}}<h<h_{c_{2}} \) when the ladder has a different ground
state, again with a simple structure.

\begin{figure}[ihtbp]
{\centering \resizebox*{5in}{!}{\includegraphics{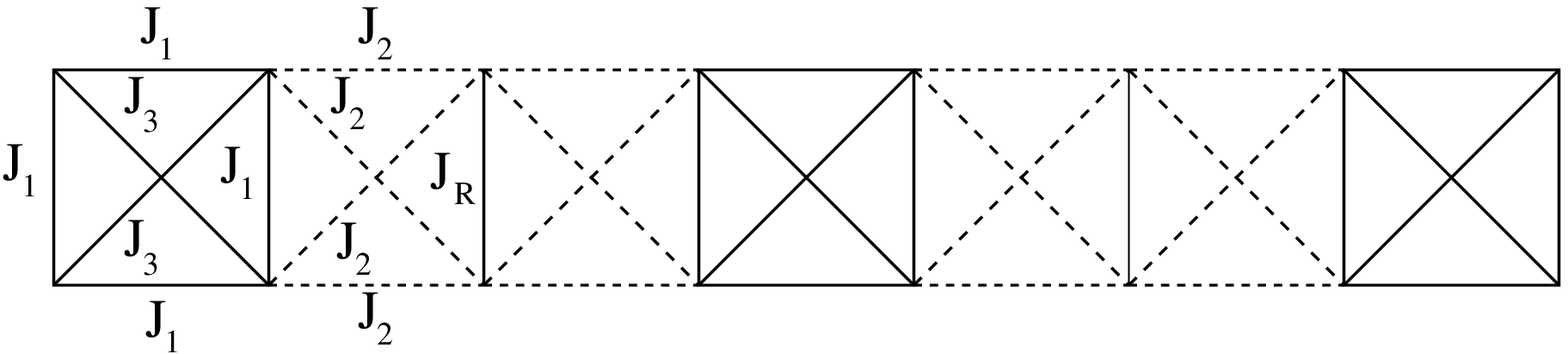}} \par}

{\centering FIG. 3: Two-chain ladder model consisting of four-spin
plaquettes coupled to two-spin rungs. The exchange\par}

{\centering interaction strengths are as shown in the Figure. \par}
\end{figure}

Fig. 3 shows another spin ladder model with modulated exchange interactions
\cite{key-21}. The model consists of four-spin plaquettes coupled
to two-spin rungs (solid lines) through NN and diagonal exchange interactions
(dotted lines) of strength \( J_{2} \). Within each plaquette, the
NN and diagonal exchange interactions are of strength \( J_{1} \)
and \( J_{3} \) respectively. The rung exchange interaction strength
is  \( J_{R} \). In a wide parameter regime, the exact ground state
has a simple product form: the plaquettes and the rungs are individually
in their ground state spin configurations. The ground state of a plaquette
is a resonating valence bond (RVB) state with wavefunction given by
\( \psi _{RVB1}(\psi _{RVB2}) \) for \( J_{3}<J_{1}(J_{3}>J_{1}) \).
The ground state of a rung is a singlet. The RVB states are given
by\\
\vspace{0.3cm} \begin{equation} \label{6} \mbox{ {\resizebox*{!}{2.0cm}{\includegraphics{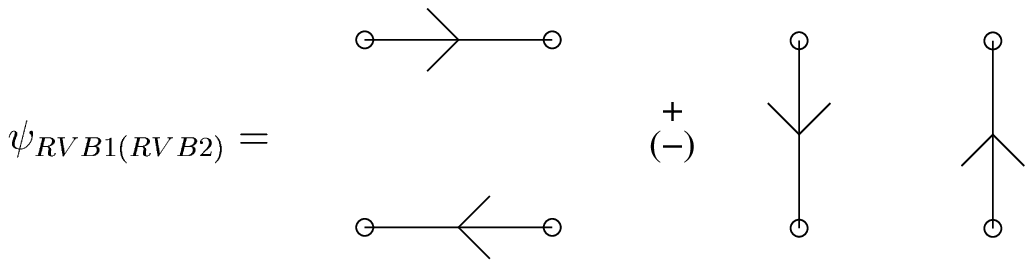}}}} \end{equation} \vspace{0.3cm}~\\
The solid lines represent singlets and the arrow signs are drawn according
to the phase convention that in a VB between the sites \( i \) and
\( j \), if the arrow points away from the site \( i \), then the
spin configuration is \( \frac{1}{\sqrt{2}}\left( \left| \uparrow (i)\downarrow (j)\right\rangle -\left| \downarrow (i)\uparrow (j)\right\rangle \right)  \).
It is easy to check that the NN spins in \( \left| \psi _{RVB1}\right\rangle  \)
are entangled with concurrence \( C=0.5 \). The NNN spins along the
diagonals are unentangled. On the other hand, the NN spins in \( \left| \psi _{RVB2}\right\rangle  \)
are unentangled and the NNN spin pairs are entangled with \( C=1 \).
At \( J_{1}=J_{3} \), there is a QPT from the ground state in which
the plaquette spin configurations are described by \( \left| \psi _{RVB1}\right\rangle  \)
to the ground state in which the same are described by \( \left| \psi _{RVB2}\right\rangle  \).
In both the phases, the rungs are in singlet spin configurations.
The QPT is accompanied by macroscopic changes in the amount of NN
and NNN entanglements. The average concurrence per NN bond, \( f_{NN} \),
in the full ladder is \( \frac{1}{3} \) for \( J_{3}<J_{1} \) and
\( \frac{1}{9} \) for \( J_{3}>J_{1} \). The average concurrence
per NNN bond, \( f_{D} \), is \( 0 \) for \( J_{3}<J_{1} \) and
\( \frac{1}{3} \) for \( J_{3}>J_{1} \). In a finite magnetic field
\( h\neq 0 \), the exact ground state maintains its product form
in an extended parameter regime. This gives rise to magnetization
plateaus in the magnetization/site \( m \) versus \( h \) curves.
Again, the jumps in the magnetization are accompanied by jumps in
the amount of entanglement. To give one specific example, consider
the case \( J_{3}<J_{1} \). The average concurrence/NNN bond in the
full ladder is zero for \( 0<h<h_{c_{1}}=J_{1} \) and \( \frac{1}{6} \)
for \( h_{c_{1}}<h<h_{c_{2}}=2J_{2} \). At \( h_{c_{1}} \), \( m \)
jumps from zero to the value \( \frac{1}{6} \). Each plaquette is
in the ground state spin configuration \( \frac{1}{2}\left( \left| \uparrow \uparrow \uparrow \downarrow \right\rangle -\left| \uparrow \uparrow \downarrow \uparrow \right\rangle -\left| \uparrow \downarrow \uparrow \uparrow \right\rangle +\left| \downarrow \uparrow \uparrow \uparrow \right\rangle \right)  \)
in the field range \( h_{c_{1}}<h<h_{c_{2}} \).
\begin{figure}[ihtbp]
{\centering \includegraphics{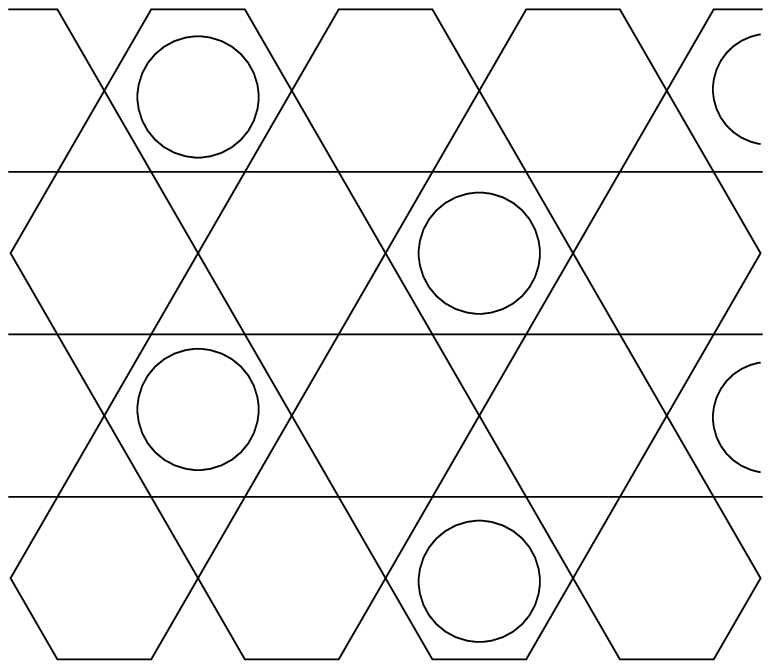} \par}

{\centering FIG. 4: Kagom\'e lattice with N spins. The circles mark
the \( \frac{N}{9} \) hexagons in which independent magnons can be
localised.\par}
\end{figure}
Schulenberg et al. \cite{key-22} have constructed exact eigenstates
which consist of independent, localized one-magnon states in a class
of frustrated spin lattices and have shown that these are the ground
states in high magnetic fields. The eigenstates are obtained provided
certain conditions are satisfied. In a kagom\'e lattice, the magnon
states are localised in \( \frac{N}{9} \) hexagons where \( N \)
is the total number of spins. The hexagons in which the magnon excitations
occur are isolated from each other (Fig. 4). Above the saturation
magnetic field, all the spins in the lattice are in a FM spin configuration
\( \left| 0\right\rangle  \) and each spin pair is unentangled. The
state \( \left| 0\right\rangle  \) is the vacuum for magnon excitations.
At a critical field below the saturation magnetic field, the exact
eigenstate consisting of \( \frac{N}{9} \) localised one-magnon excitations
becomes the ground state. The localised magnon excitation has the
wave function\begin{equation}
\label{7}
\left| 1\right\rangle =\frac{1}{\sqrt{6}}\sum ^{6}_{l=1}(-1)^{l}S^{-}_{l}\left| 0\right\rangle .
\end{equation}
This gives rise to a macroscopic change in the magnetization curve.
In each of the \( \frac{N}{9} \) hexagons, a spin is equally entangled
with all the other five spins and the magnitude of the concurrence
is \( C=\frac{1}{3}. \) This signifies a macroscopic change in the
amount of entanglement. The magnetization jump occurring at a critical
value of the magnetic field signifies a first order QPT. We have shown
through specific examples that such QPTs give rise to macroscopic
jumps in the amount of entanglement associated with NN and/or further-neighbour
spin pairs.

QPTs can also be brought about by tunning exchange interaction strengths.
We have already given an example of this in the case of the ladder
model shown in Fig. 3. For \( h=0 \), a QPT occurs at \( \frac{J_{3}}{J_{1}}=1. \)
In the frustrated two-leg ladder model shown in Fig. 1, a QPT takes
place at the critical value \( (\frac{J_{R}}{J})_{c}\simeq 1.401. \)
Below the critical point, the ground state is that of an effective
\( S=1 \) chain with the \( S=1 \) spins forming out of the pairs
of \( S=\frac{1}{2} \) rung spins. The spin ladder is now in the
Haldane phase of a \( S=1 \) chain. Above the critical point, the
ladder is in the rung-singlet (RS) phase in which the rung spins are
in singlet spin configurations. Kolezhuk and Mikeska \cite{key-23}
have proposed a generalised frustrated ladder model in which the first
order QPT between the RS and the Haldane phases can be studied in
an exact manner. This model includes biquadratic interactions besides
NN and NNN (diagonal) ones. The Hamiltonian is given by, 

\[
H=\sum _{i}h_{i,i+1},\]
\[
h_{i,i+1}=\frac{y_{1}}{2}(\overrightarrow{S}_{1,i}.\overrightarrow{S}_{2,i}+\overrightarrow{S}_{1,i+1}.\overrightarrow{S}_{2,i+1})+(\overrightarrow{S}_{1,i}.\overrightarrow{S}_{1,i+1}+\overrightarrow{S}_{2,i}.\overrightarrow{S}_{2,i+1})\]
\[
+y_{2}(\overrightarrow{S}_{1,i}.\overrightarrow{S}_{2,i+1}+\overrightarrow{S}_{2,i}.\overrightarrow{S}_{1,i+1})+x_{1}(\overrightarrow{S}_{1,i}.\overrightarrow{S}_{1,i+1})(\overrightarrow{S}_{2,i}.\overrightarrow{S}_{2,i+1})\]
\begin{equation}
\label{h1}
+x_{2}(\overrightarrow{S}_{1,i}.\overrightarrow{S}_{2,i+1})(\overrightarrow{S}_{2,i}.\overrightarrow{S}_{1,i+1})
\end{equation}
\begin{equation}
\label{h2}
\textrm{where},\textrm{ }x_{1}=\frac{4}{5}(3-2y_{2}),\textrm{ }x_{2}=\frac{4}{5}(3y_{2}-2).
\end{equation}
 The indices 1 and 2 correspond to the lower and upper legs of the
ladder respectively and \( i \) is the rung index. The exact phase
boundary between the RS and the Haldane phases is given by \( y_{1}=\frac{4}{5}(1+y_{2}). \)
The ground state in each phase is known exactly. In the Haldane phase
the ground state is the AKLT state \cite{key-18}. The ground state
energy/rung in the RS and the Haldane phases are \( E_{RS}=-\frac{3}{4}y_{1}+\frac{3}{20}(1+y_{2}) \)
and \( E_{AKLT}=\frac{1}{4}y_{1}-\frac{13}{20}(1+y_{2}) \) respectively.

As mentioned before, in the RS phase the spin pairs along the rungs
are perfectly entangled with concurrence \( C=1. \) All the other
spin pairs are unentangled. The amount of entanglement in the AKLT
phase can be computed in the following manner. For \( [H,S^{z}]=0, \)
where \( S^{z} \) is the \( z \) component of the total spin, the
reduced density matrix of a spin pair located at the sites \( i \)
and \( j \) has the form \cite{key-9}\begin{equation}
\label{ro}
\rho _{ij}=\left( \begin{array}{cccc}
u_{+} & 0 & 0 & 0\\
0 & w_{1} & z & 0\\
0 & z & w_{2} & 0\\
0 & 0 & 0 & u_{-}
\end{array}\right) .
\end{equation}
 The concurrence quantifying the entanglement is given by\begin{equation}
\label{c}
C=2\textrm{max}[0,\left| z\right| -\sqrt{u_{+}u_{-}}].
\end{equation}
 Wang and Zanardi \cite{key-24} have shown that the matrix element
of \( \rho _{ij} \) can be expressed in terms of the various correlation
functions \( G_{\alpha \beta }=\left\langle \sigma _{i\alpha }\sigma _{j\beta }\right\rangle =Tr(\sigma _{i\alpha }\sigma _{j\beta }\rho ), \)
\( (\alpha =x,y,z), \) where \( \rho  \) is the density operator.
The magnetization/site \( m=\frac{1}{N}Tr(\sum ^{N}_{i=1}\sigma _{i}^{z}\rho ). \)
In particular, the following relations hold true:\[
u_{\pm }=\frac{1}{4}(1\pm 2m+G_{zz})\]
\begin{equation}
\label{rel}
z=\frac{1}{4}(G_{xx}+G_{yy}).
\end{equation}
 The correlation functions in the AKLT phase can be calculated in
an exact manner using the transfer matrix method in the matrix product
(MP) formalism \cite{key-25}. The AKLT ground state can be written
in the MP form as \begin{equation}
\label{12}
\psi _{AKLT}=Tr\prod ^{\frac{N}{2}}_{i=1}g_{i}
\end{equation}
\begin{equation}
\label{13}
\textrm{with},\textrm{ }g_{i}=\left[ \begin{array}{cc}
\left| t_{0}\right\rangle _{i} & -\sqrt{2}\left| t_{+}\right\rangle _{i}\\
\sqrt{2}\left| t_{-}\right\rangle _{i} & -\left| t_{0}\right\rangle _{i}
\end{array}\right] .
\end{equation}
 The product in Eq. (\ref{12}) is over the \( \frac{N}{2} \) rungs
of the ladder. The two spin-\( \frac{1}{2} \)'s of each rung are
in a triplet spin configuration in the AKLT phase giving rise to an
effective spin 1. In Eq. (\ref{13}), the states \( \left| t_{\mu }\right\rangle  \)
with \( \mu =+1, \) 0 and \( -1 \) represent a spin state with \( S^{z}=+1, \)
0 and \( -1 \) respectively. Calculation of the correlation functions
\( G_{\alpha \alpha }\textrm{ }(\alpha =x,y,z) \) of the spin \( \frac{1}{2} \)
pairs in the AKLT state can be carried out following standard procedure
\cite{key-25}. We quote the final results. The various correlation
functions are:\[
\left\langle \sigma _{1,n\alpha }\sigma _{1,n+1\alpha }\right\rangle =\left\langle \sigma _{2,n\alpha }\sigma _{2,n+1\alpha }\right\rangle =\left\langle \sigma _{1,n\alpha }\sigma _{2,n+1\alpha }\right\rangle =\left\langle \sigma _{1,n+1\alpha }\sigma _{2,n\alpha }\right\rangle \]
\begin{equation}
\label{14}
=-\frac{4}{9}\textrm{ }(\alpha =x,y,z)
\end{equation}
\begin{equation}
\label{15}
\left\langle \sigma _{1,n\alpha }\sigma _{2,n\alpha }\right\rangle =\frac{1}{3}\textrm{ }(\alpha =x,y,z).
\end{equation}
 The correlation functions, in which the distance \( l \) separating
the rungs on which the spins are located is \( \geq 1 \) (in Eq.
(\ref{14}) \( l=1 \)), involve the factor \( 4(-1)^{l}3^{-l-1} \).
With the knowledge of the correlation functions, the concurrence can
be determined using the relations (\ref{c}) and (\ref{rel}) (\( m \)
in the AKLT phase is zero). One finds that in the AKLT phase, the
spin pairs along the rungs are unentangled. In the RS phase, the same
spins are maximally entangled. The intra-leg NN and the NNN (diagonal)
spin pairs are entangled in the AKLT phase with concurrence \( C=\frac{1}{6} \)
in each case. In the RS phase, these spin pairs are unentangled. Again,
the first order QPT from the RS phase to the AKLT phase is accompanied
by macroscopic changes in the entanglement structure. 

The generalised ladder model studied by Kolezhuk and Mikeska has a
rich phase diagram \cite{key-23} describing both first order and
second order QPTs. At the second order phase boundary, the gap in
the excitation spectrum goes to zero. There are five possible phases:
RS, AKLT, FM, D1 and D2. The phases D1 and D2 have spontaneously broken
symmetry but the exact ground states are not known in these phases.
The phase boundaries separating the FM-D1, AKLT-D1, AKLT-D2, RD-D1
and RD-D2 are known exactly and have been determined in the MP formalism.
The corresponding QPTs are second order transitions. The two first
order phase boundaries separating the RD-AKLT and AKLT-FM phase are
also known exactly. Preliminary calculations \cite{key-26} near the
second order phase boundaries suggest that the pairwise entanglement
does not extend beyond the NNN distance. The same is true in the case
of first order QPTs. Osterloh et al. \cite{key-13} have considered
the range of entanglement in the vicinity of the quantum critical
point of the transverse Ising model in 1D. They have found that even
at the critical point, where the spin-spin correlations are long ranged,
the concurrence is zero for spin pairs separated by more than NNN
distance. There is, however, one crucial difference between this model
and the AKLT-type models described by MP states. The MP states are
finitely correlated, i.e., the spin-spin correlation function \( \left\langle S_{i}^{z}S_{i+l}^{z}\right\rangle =A_{S}e^{-\frac{l}{\xi }} \)
decays exponentially with the correlation length \( \xi  \) equal
to a few lattice spacings. As the transition point \( \tau =\tau _{c} \)
(\( \tau  \) is some model parameter) is approached, the correlation
length \( \xi  \) either does not exhibit any singularity or diverges
in a power-law fashion as \( (\tau -\tau _{c})^{2} \). In the latter
case, however, the prefactor \( A_{s}\propto (\tau -\tau _{c}) \)
becomes zero at the transition point \cite{key-23}. Thus long range
spin-spin correlations cannot develop in the system.

In summary, we have shown through specific examples that first order
QPTs can bring about macroscopic changes in the amount of pairwise
entanglement in spin systems. We have given the specific example of
a spin model in which magnetization plateaus give rise to a plateau
structure in the average concurrence per rung as well as NN bond.
There are several low-dimensional AFM compounds in which magnetization
plateaus have been observed experimentally \cite{key-27}. The appearance
of magnetization plateaus has been explained in terms of metal-insulator
transitions of magnetic excitations driven by a magnetic field \cite{key-32}.
In the insulating (plateau) phase, the magnetic excitations give rise
to crystalline order and in the metallic (non-plateau) phase they
are itinerant. It will be of interest to explore the possibility of
a plateau structure in the amount of \( T=0 \) as well as finite
\( T \) entanglements in such systems. Our study shows that an external
magnetic field can be employed to give rise to large changes in the
amount of entanglement and provides the basics for the construction
of an entanglement `amplifier' or `switch'. There is a large number
of spin models which exhibit QPTs of significant interest. Some of
these models describe 2D systems. Examples include the Shastry-Sutherland
model \cite{key-28}, the \( S=\frac{1}{2} \) AFM model on the \( \frac{1}{5} \)-depleted
square lattice \cite{key-29}, a lattice of weakly-coupled two-leg
ladders \cite{key-30} etc. Some of these models are reviewed in Refs.
\cite{bose,key-31}. All these models exhibit second order QPTs from
a spin-disordered gapped phase to a gapless phase with long range
spin-spin correlations. Osterloh et al. \cite{key-13} and Osborne
and Nielsen \cite{key-14} have found evidence of entanglement showing
scaling behaviour in the vicinity of the quantum critical point. Similar
studies should be carried out for the spin models mentioned to augment
our knowledge of the relationship between QPTs and entanglement. 

\textbf{Acknowledgement}

E. Chattopadhyay is supported by the Council of Scientific and Industrial
Research, India under sanction No. 9/15(186)/97-EMR-I.

\end{document}